# On the theory of phase transition in muliferroics

**B R Gadjiev**
International University for Nature, Society and Man, 19 Universitetskaya Street, Dubna, 141980, Russia

E-mail: gadjiev@uni-dubna.ru

**Abstract.** In this paper we investigate peculiarities of phase transition high-symmetry - incommensurate phase in inhomogeneous systems. We obtain the nonlinear dispersion law and then present a renormalization group analysis of phase transitions in multiferroics. We have determined the dependence of critical indices on the nonextensivity parameter of the system.

In the magnetoelectric multiferroic materials magnetic and ferroelectric order coexists and mutually interacts [1]. Classification of multiferroic materials on the level of microscopic mechanisms is presented in review [2], where analyses how multiferroics combine the properties of ferroelectrics and magnets have been done. In the multiferroic materials near the phase transition we deal with non local interactions in the system with long range correlations. In this paper we investigate peculiarities of phase transition high-symmetry - incommensurate phase in such inhomogeneous systems.

Let us assume that a crystal with the average space group $G$ in the paramagnetic phase undergoes a magnetic phase transition at critical point $T_c$. Let the order parameter ($\eta_1$, $\eta_2$) (magnetization components) transformed by the active irreducible representation $D$ of the symmetry group of the paramagnetic phase corresponding to point $\boldsymbol{q_0} = (0,0,0)$ of Brillouin zone. If the compound in the paramagnetic phase has rhombohedral symmetry, the symmetry analysis shows that the following gradient invariants are possible: $gP_z\left(\eta_1 \frac{\partial \eta_2}{\partial y} - \eta_2 \frac{\partial \eta_1}{\partial y}\right)$ and $\sigma P_y\left(\eta_1 \frac{\partial \eta_2}{\partial z} - \eta_2 \frac{\partial \eta_1}{\partial z}\right)$, where $P_z$ and $P_y$ are components of the polarization vector. In this case, the symmetry analysis shows that the density of the free energy functional after introducing the variables $\eta_1 = \rho \sin \varphi(y, z)$ and $\eta_2 = \rho \cos \varphi(y, z)$, and using the constant amplitude approximation contains the following integer basis of invariants

$$\rho^2, \rho^n \cos n\varphi, (P_y^2 + P_z^2), \rho^{n/2}(P_z + P_x)\sin \frac{n}{2}\varphi \qquad (1)$$

Thus, in one dimensional case the free energy functional with non local interaction can be presented as $F[\varphi, P] = F_0[\varphi, P] + F_1[\varphi, P]$, where

$$F_0[\varphi, P] = \int dz \int dx \left[\frac{k}{2}\rho^2 \frac{\partial \varphi(z)}{\partial z} g_1(z, x) \frac{\partial \varphi(x)}{\partial x} - \sigma \rho^2 P(z) g(z, x) \frac{\partial \varphi(x)}{\partial x} + \xi \rho^{n/2} P(z) g_0(z, x) \sin \frac{n}{2}\varphi(x)\right] \qquad (2)$$

and

$$F_1[\varphi, P] = \int dx \left[ \frac{\alpha}{2} \rho^2(x) + \frac{\beta}{4} \rho^4(x) + \gamma \rho^n(x) \cos n\varphi(x) + \frac{\chi}{2} P^2(x) \right] \qquad (3)$$

The equations of motion for the order parameters will be derived by using the Gateaux differential of $F[\varphi, P]$ at the points $\varphi(x)$ and $P(x)$, which is defined as the limit $\delta F[\eta(x)] = \lim_{\varepsilon \to 0}(1/\varepsilon)(F[\eta(x) + \varepsilon \mu(x)] - F[\eta(x)])$. Equation of motion for the phase $\varphi(x)$ and the polarization $P(x)$ is defined from condition $\delta F = 0$ and has the form

$$\int dz \left[ -\frac{k}{2} \rho^2 \frac{\partial k_1(x,z)}{\partial x} \frac{\partial \varphi(z)}{\partial z} + \sigma \rho^2 P(z) \frac{\partial g(z,x)}{\partial x} + \frac{n}{2} \xi \rho^{n/2} P(z) g_0(z,x) \cos \frac{n}{2} \varphi(x) \right] + n\gamma \rho^n \sin n\varphi = 0, \qquad (4)$$

$$\chi P(z) = \int dx \left[ -\sigma \rho^2 g(z,x) \frac{\partial \varphi(x)}{\partial x} + \xi \rho^{n/2} \sin \frac{n}{2} \varphi(x) g_0(z,x) \right]. \qquad (5)$$

Let's consider a case of local interaction in space, a namely $k_1(x,z) = k_1^0 \delta(x-z)$ and $g_0(x,z) = g_0^0 \delta(x-z)$ and $g(x,z) = g^0 \delta(x-z)$. In this case

$$\frac{c}{2} \frac{\partial^2 \varphi(x)}{\partial x^2} + \sigma \rho^2 g^0 \frac{\partial P(x)}{\partial x} + \frac{n}{2} \xi \rho^{n/2} P(x) \cos \frac{n}{2} \varphi(x) + n\gamma \rho^n \sin n\varphi(x) = 0, \qquad (6)$$

$$\chi P(x) = -\sigma \rho^2 g^0 \frac{\partial \varphi(x)}{\partial x} + \xi g_0^0 \rho^{n/2} \sin \frac{n}{2} \varphi(z). \qquad (7)$$

Let's assume, that $\xi = 0$ and we obtain

$$\frac{c}{2} \frac{\partial^2 \varphi(x)}{\partial x^2} + \sigma \rho^2 g^0 \frac{\partial P(x)}{\partial x} + n\gamma \rho^n \sin n\varphi(x) = 0, \qquad (8)$$

$$\chi P(z) = -\sigma \rho^2 g^0 \frac{\partial \varphi(x)}{\partial x}. \qquad (9)$$

After substitution equation (9) in the equation (8) we obtain

$$\bar{c} \frac{\partial^2 \varphi(x)}{\partial x^2} + n\gamma \rho^n \sin n\varphi(x) = 0, \qquad (10)$$

Solution of equation (10) has the form

$$\varphi(x) = \frac{2}{n} am(x). \qquad (11)$$

In a sinusoidal regime of an incommensurate phase $\varphi(x) = \kappa x$, where $\kappa$ is a wave vector of structure. Thus, the incommensurate phase is represented in the form of alternating domains with the opposite magnetic moments. If $\xi \neq 0$ then the equation (6) contains the first derivative of a phase and therefore, the amplitude of a phase decreases with distance.

At presence of non local interaction the incommensurate phase is described by the following equation

$$\int dz \left[ -\frac{k}{2} \rho^2 \frac{\partial k_1(x,z)}{\partial x} \frac{\partial \varphi(z)}{\partial z} \right] + n\gamma \rho^n \sin n\varphi = 0, \qquad (12)$$

Then the integral in equation (12) can be considered as an average on distribution $\frac{\partial k_1(x,z)}{\partial x} = P(x,y)$. In the system with long range correlations it is natural to assume that the structure possesses small world property [3]. The topology of small world network is described by $q$-exponential distribution $P(r) = Z^{-1} \exp_q(-\varsigma r)^q$, where $\exp_q(x)$ expresses the $q$-exponential function defined by $\exp_q(x) = (1+(1-q)x)^{\frac{1}{1-q}}$. It easy to see that $P(r)$ reduces to $\exp(-\varsigma r)$ in the limit $q=1$. In case of an exponential distribution $P(r)$ the equation (12) is reduced to the equation (10) with renormalized parameters. We note from $P(r) = Z^{-1}\exp_q(-\varsigma r)^q$ that $P(r) \sim r^{-\frac{q}{q-1}}$, where $q > 1$. In this case the equation (12) becomes

$$a\frac{\partial^\alpha \varphi_0(z)}{\partial r^\alpha} + \sin\varphi_0(z) = 0, \qquad (13)$$

where $\varphi_0(z) = n\varphi(z)$ and $1 < \alpha \leq 2$. In sinusoidal regime an incommensurate phase is described by solution $\eta \sim \sin\kappa x$. Analysis of equation (13) shows that $\eta \sim \sin\left(\frac{(kx)^{\alpha-1}}{\Gamma(\alpha)}\right)$. Thus, in this case distances between arbitrary nearest neighbors zero of the phase of order parameter behaves as $\sim \frac{1}{\alpha-1}[\pi\Gamma(\alpha)m]^{\frac{1}{\alpha-1}}$ and, therefore, increases when number of zero increases. Besides, this distance increases when decreases $\alpha$.

The equation of motion for the order parameter is

$$\int_0^t dt' P_0(t-t')\frac{\partial \eta(x,t')}{\partial t'} + \int_{-\infty}^{+\infty} dy P_1(x-y)\frac{\partial \eta(y,t)}{\partial y} + \frac{\partial F(\eta(x,t))}{\partial \eta(x,t)} = 0, \qquad (14)$$

where $P_0(t)$ and $P_1(x)$ are Tsallis distribution, $F(\eta(x,t))$ and $\eta(x,t)$ are a free energy and a order parameter, accordingly. This equation leads to the nonlinear dispersion law, i.e. $\omega(q) = \alpha + q^{1+\nu}$ where $\alpha$ is reduced temperature.

For the renormalization group analysis of high symmetry–incommensurate phase transition we consider [4]

$$Z = \int\left(\prod_{\vec{k}} d^n \sigma_{\vec{k}}\right)\exp_q(H[\sigma_{\vec{k}}]), \qquad (15)$$

If the symmetry group of the paramagnetic phase contains symmetry element of space inversion, the effective Hamiltonian, necessary for the renormalization group analysis, has the form:

$$H[\sigma(\vec{q})] = -\frac{1}{2}\int_{\vec{q}} \omega(\vec{q})\sigma(\vec{q})\sigma(-\vec{q}) - \int_{\vec{q}}\int_{\vec{q}'}\int_{\vec{q}''} \sigma(\vec{q})\sigma(\vec{q}')\sigma(\vec{q}'') \qquad (16)$$

where $\int_{\vec{k}} = \frac{1}{\Omega_0}\int d^d k$. The integral over $\vec{k}$ is over the appropriate Brillouin zone.

Assuming the presence of the fixed point, we obtain the renormalization group equations:

$$r' = \xi^2 b^{-d}[r + 4(n+2)uA(r) + ...], \qquad (17)$$

$$u' = b^{-3d}\xi^4\left[u - 4(n+8)u^2 C(r) + ...\right]. \tag{18}$$

We have introduced the following notations:

$$A(r) = \int_{\vec{q}} \omega(\vec{q}), \quad C(r) = \int_{\vec{q}} \omega^{-2}(\vec{q}), \tag{19}$$

where integration in the momenta is produced in the region $\Lambda/b < |\vec{q}| < \Lambda$ in the $d$−dimension space. The parameter $\xi$ is derived from the condition of equality to unit of the coefficient at $\vec{q}$ in $\omega(q)$. It produces $\xi^2 b^{-d-\gamma} = 1$, from which it follows that $\xi^2 b^{-d} = b^\gamma$, where $\gamma = 1 + \nu$. Consequently $\xi^4 b^{-3d} = b^{-2\gamma+d}$. Thus, the renormalization group equation has the form:

$$r' = b^\gamma\left[r + 4(n+2)uA(r) + ...\right], \quad u' = b^{-d+2\gamma}\xi^4\left[u - 4(n+8)u^2 C(r) + ...\right]. \tag{20}$$

Let us assume that $\gamma = \frac{3}{2} + \varepsilon$, where $\varepsilon \in (0, 1/2)$. Let us determine the critical indices in space with dimension $d = 3$. In this case, we obtain from equation (20)

$$r' = b^{(3/2+\varepsilon)}\left[r + 4(n+2)uA(r) + ...\right], \quad u' = b^{2\varepsilon}\left[u - 4(n+8)u^2 C(r) + ...\right]. \tag{21}$$

Direct calculation shows that $C(r) \approx K_d \ln b$, and in the fixed point $u^* = \dfrac{2\varepsilon}{4(n+8)K_d}$.

Taking into account that $A(r) \approx -rC(0)$, we obtain

$$r' = b^{(3/2+\varepsilon)}\left[r - 4(n+2)ruC(0) + ...\right]. \tag{22}$$

The linearization of the equation at the fixed point $u^*$ gives $\Delta r' = b^{1/\nu}\Delta r$ that allows the determination of the critical index $\nu = \dfrac{3}{2}\left(1 - \left(\dfrac{2(n+2)}{n+8}\right)\varepsilon\right)$.

where $n$ is a number of components of the order parameter. According to the similarity relation, the critical index of the magnetic susceptibility is $\gamma = 2\nu$.

In our approach the structure fractality is introduced by the function $P(x - x')$ which is characterized by the interaction of the structure defects on the dynamics of the order parameter and changes space distribution of order parameter. In the arisen phase, space distribution of the order parameter is described by aperiodic function. Besides, the determination of the function $P(x - x')$ in various fractal dimensions leads to various values of the order of the fractional differential equation that describes the motion of the order parameter. It changes the relaxation law of the order parameter and leads to the nonlinear dispersion law. The renormalization group analysis shows that in this case a new critical regime appears, together with the dependence of critical values on the structure fractal dimension.